\documentclass[journal]{IEEEtran}
\usepackage{cite}

\ifCLASSINFOpdf 
\usepackage[pdftex]{graphicx}
\else
\usepackage[dvips]{graphicx}
\fi

\usepackage[cmex10]{amsmath}
\usepackage{algorithmic}
\usepackage[ruled,norelsize]{algorithm2e}

\usepackage{array}

\usepackage{mdwmath}
\usepackage{mdwtab}
\usepackage{todonotes}

\usepackage{eqparbox}

\ifCLASSOPTIONcompsoc
\usepackage[caption=false,font=normalsize,labelfont=sf,textfont=sf]{subfig}
\else
\usepackage[caption=false,font=footnotesize]{subfig}
\fi

\usepackage{fixltx2e}
\usepackage{url}
\usepackage{amsfonts}
\usepackage{amssymb}
\usepackage{amsthm}
\usepackage{bm}
\usepackage[ruled]{algorithm2e}
\usepackage[utf8]{inputenc}
\usepackage{booktabs}
\usepackage{url}
\usepackage{cite}
\usepackage{flushend}
\usepackage{caption}
\usepackage{makecell}
\usepackage{url}
\usepackage{epstopdf}

\newcounter{tempEquationCounter} 
\newcounter{thisEquationNumber}

\begin{document}
%
\title{Toward Quantum-Safe 6G: Experimental Evaluation of Post-Quantum Cryptography Techniques}


\author{Ananya Kudaloor, Adnan Aijaz,~\IEEEmembership{Senior Member,~IEEE}
\vspace{-0.4cm}
\thanks{The authors are with the Bristol Research and Innovation Laboratory, Toshiba Europe Ltd., Bristol, BS1 4ND, UK. Contact e-mail: firstname.lastname@toshiba.eu}}
\markboth{IEEE Communications Standards Magazine -- Accepted for Publication}%
{Shell \MakeLowercase{\textit{et al.}}: Bare Demo of IEEEtran.cls for Journals}
%


\maketitle
\begin{abstract}
\boldmath
6G networks will require quantum-secure cryptography deployed across core infrastructure, edge nodes, resource-constrained IoT devices. Although post-quantum cryptographic (PQC) algorithms have been standardized by NIST, their practical deployability in bandwidth and latency limited wireless systems remains unclear. This paper presents a practical evaluation of NIST selected PQC schemes, including ML-KEM (Kyber), ML-DSA (Dilithium), and Falcon. Benchmarks conducted with OpenSSL and the OQS provider on heterogeneous platforms show that while computational performance is acceptable, ciphertext and signature size expansion significantly impact handshake reliability and bandwidth efficiency, particularly at the network edge. The results highlight key system-level trade-offs and motivate the need for PQC optimization and deployment-aware design for future quantum-secure 6G networks.

\end{abstract}



%
\IEEEpeerreviewmaketitle

\section{Introduction}
\IEEEPARstart {6}{G} networks are envisaged to create the Internet-of-Everything -- a hyper-connected world of people and machines, and physical and digital realities \cite{saad_6g_vision}. At the same time, advances in quantum computing pose a fundamental threat to the public-key cryptography that underpins today’s Internet security. Shor’s algorithm demonstrates that sufficiently large, fault-tolerant quantum computers requiring millions of physical qubits and thousands of logical qubits could efficiently break widely deployed schemes such as RSA and elliptic curve cryptography (ECC), compromising confidentiality, authentication, and long-term data security.

Looking ahead, large scale quantum computers pose a fundamental risk to today’s encrypted communications: data protected using classical public-key cryptography can be harvested now and retrospectively decrypted once quantum attacks become practical. This “harvest now, decrypt later” threat has accelerated the transition toward post-quantum cryptography (PQC), which aims to provide resistance against both classical and quantum adversaries. Following the finalization of NIST’s PQC standards in 2024 selecting ML-KEM for key establishment and ML-DSA for digital signatures major Internet infrastructure providers have begun deploying PQC in production environments.

At Internet scale, Cloudflare has played a significant role in operationalizing PQC by deploying hybrid key agreement mechanisms combining ML-KEM with classical X25519 to secure TLS 1.3 traffic across its edge and internal systems \cite{cloudflare_automatically_secure}. As of late 2025, a substantial fraction of human generated connections to Cloudflare are already protected using hybrid post-quantum key exchange, demonstrating the feasibility of PQC deployment at scale \cite{cloudflare_automatically_secure}. However, extending post-quantum protection beyond the edge particularly toward origin servers introduces additional complexity. Origin servers operate in a highly heterogeneous ecosystem of software stacks, middleboxes, and legacy configurations, many of which do not yet support hybrid post-quantum handshakes.

Complementary deployments by cloud providers further highlight the system-level implications of PQC adoption. Google Cloud has introduced quantum-safe key encapsulation mechanisms within Cloud Key Management Service (KMS), enabling hybrid-classical post-quantum key establishment for managing cryptographic workflows \cite{googlecloud_pqc_kms}. Amazon Web Services (AWS) has similarly deployed hybrid ML-KEM-based post-quantum TLS for critical services such as AWS KMS, Certificate Manager (ACM), and Secrets Manager, and has quantified the associated performance overheads \cite{aws_mlkem_pqtls}. Specifically, migrating from an elliptic-curve Diffie–Hellman (ECDH) key agreement to an ECDH + ML-KEM hybrid requires transmitting approximately 1600 additional bytes during the TLS handshake and incurs an additional 80–150 µs of cryptographic computation for ML-KEM operations \cite{aws_mlkem_pqtls}. Importantly, this overhead is incurred only during connection establishment and is amortized over the lifetime of the TLS session across subsequent application layer requests.

These deployments show that PQC is viable at Internet scale, but also reveal that latency, bandwidth expansion, and energy use can limit feasibility in edge and IoT settings. While NIST standardization ensures cryptographic soundness, it does not address system-level trade-offs in real protocols and heterogeneous hardware, making empirical evaluation crucial for future 6G deployment decisions.

\subsection{Contributions}
This paper presents a deployment-oriented evaluation of NIST-standardized PQC schemes for 6G, covering computational cost, payload growth, ciphertext expansion rate, and security implications. The main contributions are:
\begin{itemize}
    \item An evaluation on primitive performance, TLS~1.3 handshake behavior, and transport-layer payload measurements.
    \item Quantitative analysis of ciphertext expansion and its propagation to signaling overhead.
    \item Comparative assessment across high-performance and constrained platforms, revealing deployment trade-offs for core and IoT scenarios.
    \item Discussion of reliability, interoperability, and standardization considerations for secure PQC integration in 6G architectures.
\end{itemize}


\begin{figure*}[t]
    \centering

    \begin{minipage}{0.45\textwidth}
        \centering
        \includegraphics[width=\linewidth]{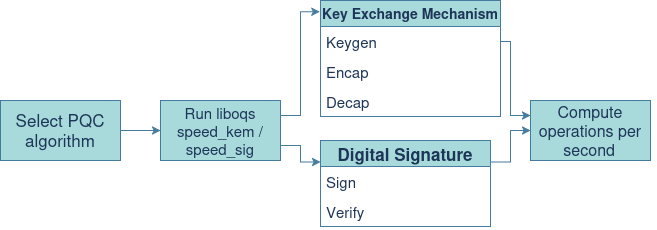}\\
        \small{(a)}
    \end{minipage}
    \hfill
    \begin{minipage}{0.45\textwidth}
        \centering
        \includegraphics[width=\linewidth]{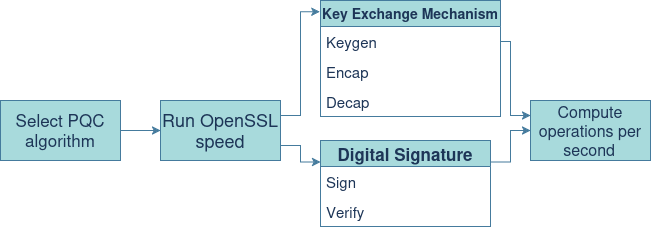}\\
        \small{(b)}
    \end{minipage}

    \vspace{2mm}

    \begin{minipage}{0.47\textwidth}
        \centering
        \includegraphics[width=\linewidth]{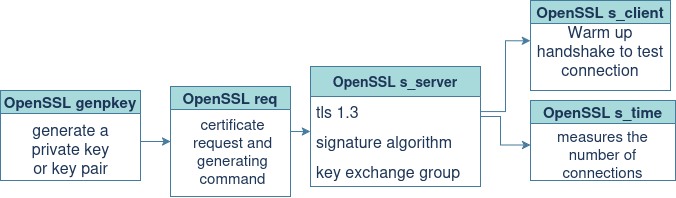}\\
        \small{(c)}
    \end{minipage}
    \hfill
    \begin{minipage}{0.47\textwidth}
        \centering
        \includegraphics[width=\linewidth]{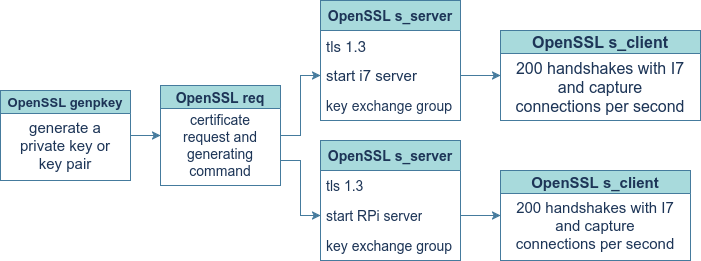}\\
        \small{(d)} 
    \end{minipage}

    \caption{End-to-end evaluation workflows for PQC. 
    (A) Algorithm level evaluation workflow for PQC primitives using \texttt{liboqs speed kem / speed sig} results for capturing keys generated, encapsulation and decapsulations per second.
    (B) Algorithm level evaluation workflow for post-quantum cryptographic primitives using \texttt{openssl speed} results for keys generated, encapsulation and decapsulations per second. 
    (C) TLS~1.3 handshake throughput measurement using \texttt{openssl s\_time}. 
    (D) TLS handshakes per second measurement using \texttt{openssl s\_client} across Intel~i7 and Raspberry~Pi platforms.}
    \label{fig:1}
\end{figure*}

\section{PQC Landscape for 6G}
Future 6G security architectures must support quantum-safe cryptography across highly heterogeneous environments, ranging from centralized cloud cores to latency-sensitive edge platforms and resource-constrained IoT devices \cite{sharma_qcs6g_2025}. Accordingly, the PQC landscape for 6G is shaped by standardized algorithms and protocols already being adopted in Internet and cloud infrastructures, which are expected to form the basis of future telecom security stacks.

While large-scale 6G networks are still under development, many of the core security mechanisms expected in 6G will evolve directly from 5G service-based architectures \cite{scalise_pqc_5gcore}. As a result, existing open-source 5G core implementations provide a practical platform for exploring the feasibility of PQC integration. Open-source cores such as Open5GS and Free5GC enable experimentation with cryptographic protection of control-plane and service-based interfaces that are expected to persist in future 6G designs. This enables PQC deployment challenges in both core and edge interfaces to be studied ahead of full 6G standardization.

Recent work analyzing core networks in Free5GC with PQC key encapsulation mechanisms evaluates the integration of post-quantum cryptographic schemes into 5G core network security, examining their impact on control-plane communication and protocol overhead \cite{scalise_pqc_5gcore}. These studies demonstrate that 5G core platforms can serve as realistic testbeds for assessing PQC readiness in future 6G architectures.

Beyond 5G-centric platforms, emerging research-oriented cores such as Open6GCore aim to provide a modular and extensible foundation for 6G experimentation. These platforms build upon 5G service-based principles while enabling early validation of new protocols and security mechanisms \cite{buhr_open6gcore_2025}. Together, such platforms allow PQC deployment challenges in core-to-core and edge-facing interfaces to be studied in realistic system environments before the availability of fully standardized 6G infrastructures.

\section{Related Work}

Prior work on post-quantum performance evaluation can be broadly divided into primitive-level benchmarking and protocol-level integration studies. Nagy \emph{et al.}~\cite{nagy_mlkempaper} evaluated ML-KEM at the primitive level, reporting key generation, encapsulation, and decapsulation performance across different platforms. This provides a useful baseline for algorithmic performance, but does not focus on end-to-end TLS behavior. Sim \emph{et al.}~\cite{sim_kpqc_tls} integrated and benchmarked PQC within TLS/X.509 workflows, including certificate size and handshake latency in practical software stacks. However, there is still a need to study how ciphertext expansion and the resulting TLS payload growth affect communication scalability. In addition, protocol-level benchmarking of these algorithms should consider broader system-level conditions, including energy, network impairment, and resource utilization.

\section{Evaluation Methodology}

To assess how PQC impacts future 6G deployments beyond algorithmic security guarantees, the paper adopts a deployment oriented evaluation methodology that links cryptographic performance with protocol behavior and communication overhead. The evaluation progresses from primitive-level benchmarking to protocol-level measurements, and is implemented using the Open Quantum Safe (OQS) project, which provides production-ready PQC implementations through liboqs (a C library) and an OpenSSL provider, enabling evaluation within standard TLS workflows~\cite{oqs_project}.



Standardized PQC algorithms are selected and evaluated using liboqs and OpenSSL~3.5.2 together with the Open Quantum Safe (OQS) provider 0.10.0.

\subsection{Algorithm-Level Cryptographic Primitive Evaluation}

The first stage focuses on algorithm-level evaluation of PQC primitives, isolating cryptographic performance from protocol and network effects. As illustrated in Fig.~\ref{fig:1}a, a PQC algorithm is first selected and evaluated using the liboqs benchmarking utilities.

For key encapsulation mechanisms, the \texttt{speed\_kem} tool repeatedly executes key generation, encapsulation, and decapsulation operations, similar to the results from Nagy \emph{et al.}~\cite{nagy_mlkempaper}. For digital signature schemes, the \texttt{speed\_sig} tool measures signing and verification operations. The output of these benchmarks reports throughput in operations per second, providing a direct measure of computational cost under identical software and hardware conditions.

In addition to liboqs, primitive-level benchmarks are also performed using the \texttt{openssl speed} command, as shown in Fig.~\ref{fig:1}b. This workflow measures key exchange and signature operations through OpenSSL’s API, enabling direct comparison between classical and post-quantum algorithms within the same cryptographic framework. Together, these measurements establish a baseline understanding of PQC performance across different implementations.

\subsection{Protocol-Level Evaluation Using TLS~1.3} 

The second stage evaluates PQC within the TLS~1.3 protocol, capturing protocol behavior that cannot be observed through primitive benchmarks alone. This includes handshake execution, message processing, and interaction between cryptographic algorithms and protocol logic. As TLS is widely deployed in current 5G service-based architectures and 6G, it provides a practical and realistic framework to assess PQC integration at the protocol layer. All measurements correspond to full TLS 1.3 handshakes without session resumption, ensuring that each connection establishment includes complete key exchange and certificate verification steps.

As shown in Fig.~\ref{fig:1}c, PQC private keys or key pairs are first generated using \texttt{openssl genpkey}, followed by certificate generation using \texttt{openssl req}. A TLS~1.3 server is then initiated using \texttt{openssl s\_server}, explicitly configured with a chosen signature algorithm and key exchange group. Classical, PQC-only, and hybrid configurations are evaluated by modifying the TLS parameters accordingly. 

TLS clients initiate handshakes using \texttt{openssl s\_client}. To ensure stable measurements, initial warm-up handshakes are performed before data collection. 

To automate handshake throughput measurement, the \texttt{openssl s\_time} utility is used, as illustrated in Fig.~\ref{fig:1}c. This tool repeatedly establishes TLS connections over a fixed interval and reports the average number of handshakes per second. Measurements are performed on both high performance and resource-constrained platforms to capture the impact of hardware heterogeneity on PQC enabled TLS. 

Handshake performance is also quantified by repeatedly measuring the number of successful \texttt{openssl s\_client} connections completed per second as represented in Fig.~\ref{fig:1}d, following the directions of Sim \emph{et al.}~\cite{sim_kpqc_tls}. 

This stage captures the effect of PQC on secure session establishment, which is a latency sensitive operation in dynamic 6G environments involving frequent mobility events, re-authentication, and edge-assisted services.  

In addition, the same protocol-level framework has been extended to evaluate broader system-level conditions, including energy, network impairment, and resource utilization.

\subsection{Communication Overhead and Ciphertext Size Analysis}   

Beyond computational performance, communication overhead is evaluated as a key constraint for PQC deployment in bandwidth- and latency-sensitive 6G environments. Ciphertext expansion rate (CER) is computed to quantify algorithm-level growth in transmitted data. For compressed ML-KEM ciphertexts, the transmitted bit length is given by $B_{ct} = k \cdot n \cdot d_u + n \cdot d_v$, and the effective expansion rate is defined as $CER = B_{ct}/K$, where $k$, $d_u$, and $d_v$ denote the security and compression parameters, and $K$ represents the encapsulated information bits \cite{liu_sc_kyber_2025}.
To evaluate how this expansion propagates to the protocol layer, the total TCP payload bytes exchanged during a single TLS~1.3 handshake are measured for each configuration. Controlled client--server sessions are established using OpenSSL with the OQS provider enabled, and packet traces are captured. The effective handshake payload is computed by summing the \texttt{tcp.len} field of all packets carrying data during connection establishment, thereby isolating transport-layer signaling overhead per secure session.

\subsection{Summary of Methodological Scope}

This work presents a deployment-oriented framework that connects lattice-level ciphertext expansion, TLS behaviour, and transport-layer signalling overhead for future 6G architectures. The methodology combines primitive benchmarking, CER analysis from ML-KEM compression parameters, and measured TCP payload bytes per TLS handshake. Building on prior studies on primitive-level ML-KEM benchmarking~\cite{nagy_mlkempaper} and TLS/X.509 integration~\cite{sim_kpqc_tls}, this work examines how ciphertext expansion propagates to handshake payload growth and signalling load, providing a cross-layer view of communication scalability. The framework can also be extended with broader system-level factors such as energy, artificial delay, packet loss, memory utilization, and CPU utilization.



\begin{figure*}[t]
    \centering
    \includegraphics[width=15.5cm]{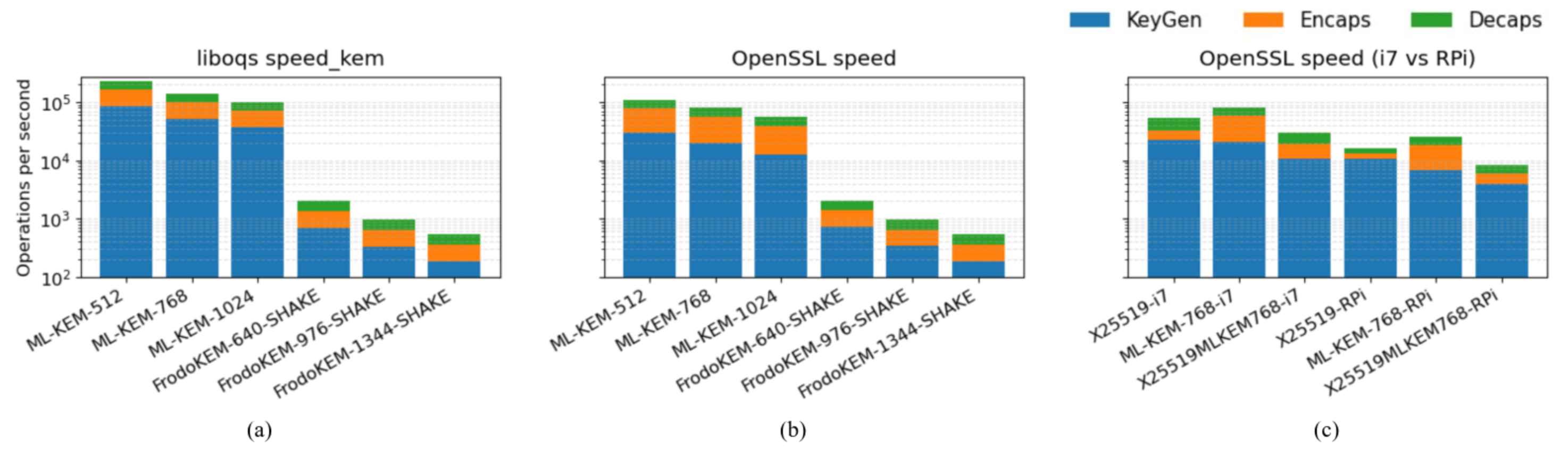}
    \caption{Primitive level performance evaluation of post-quantum key exchange schemes.
    (a) Liboqs benchmarking results,
    (b) Results of \texttt{openssl speed} evaluation
    and (c) Cross-platform comparison between Intel~i7 and Raspberry~Pi.}
    \label{fig:3}
\end{figure*}

\section{Evaluation Results}
\subsection{Cryptographic Primitive Performance Results}

This stage presents the performance results of PQC primitives evaluated using the liboqs framework on an Intel Core i7-based platform in Fig.~\ref{fig:3}a which is evaluated based on Fig.~\ref{fig:1}a. The objective of this stage is to establish a baseline comparison of standardized key encapsulation mechanisms under identical software and hardware conditions, independent of protocol or network effects.

The performance variation observed across the evaluated KEMs is closely tied to both ciphertext size and the underlying computational structure of the schemes. For ML-KEM, throughput decreases as the security level increases from ML-KEM-512 to ML-KEM-1024, reflecting the larger polynomial dimensions, increased number of arithmetic operations, and higher memory access requirements at higher parameter sets. Despite this trend, ML-KEM maintains comparatively high throughput across all security levels due to its compact ciphertexts and efficient module lattice arithmetic optimized for general purpose CPUs Fig.~\ref{fig:3}a.
In contrast, FrodoKEM-SHAKE variants exhibit lower throughput due to larger ciphertext sizes and unstructured lattice-based matrix operations. Throughput decreases further from FrodoKEM-640-SHAKE to FrodoKEM-1344-SHAKE as matrix dimensions grow. However, lower throughput does not imply weaker performance FrodoKEM relies on standard LWE without ring structure, offering conservative security assumptions and strong robustness for high-assurance deployments.

To complement the liboqs based evaluation, additional benchmarks were conducted using the \texttt{openssl speed} utility to assess primitive level performance across different hardware platforms, represented in Fig.~\ref{fig:1}b. Measurements were performed on an Intel Core i7 workstation and a Raspberry~Pi~4, enabling comparison between a high-performance desktop CPU and an ARM-class constrained reference platform. Fig.~\ref{fig:3}b represents measurements over various Key exchange algorithms.
The results in Fig.~\ref{fig:3}c indicate that the Intel i7 consistently achieves approximately two times higher key generation throughput than the Raspberry~Pi~4 across both classical and post-quantum key exchange mechanisms. This performance gap highlights the strong influence of CPU microarchitecture, instruction set extensions, memory hierarchy, and compiler level optimizations on cryptographic workloads. In particular, the i7 platform benefits from wider vector units and more aggressive optimization paths, which are less available on ARM based embedded processors.

On the Intel i7 platform, ML-KEM-512 and ML-KEM-768 exhibit key generation throughput that is comparable to classical X25519, indicating that lattice based post-quantum key exchange can approach the performance of widely deployed elliptic curve mechanisms on modern CPUs. In contrast, on the Raspberry~Pi~4, classical X25519 keys generated throughput maintains a clear performance advantage over ML-KEM variants, reflecting the lower computational complexity and more mature optimizations of classical elliptic curve arithmetic on constrained hardware Fig.~\ref{fig:3}c. These results suggest that while post-quantum schemes can be competitive on high performance platforms, hardware capabilities and ciphertext sizes play a decisive role in determining their suitability for IoT deployments. 

\subsection{TLS Handshake Performance Results}

In 6G architectures, cryptographic mechanisms are realized through network security protocols rather than isolated primitives. TLS is therefore used as the evaluation vehicle, as it underpins secure communication in current and emerging cloud-native networks. Evaluating PQC within TLS captures its impact on handshake overhead and connection setup in realistic deployments.

\begin{figure}[t]
    \centering
    \includegraphics[width=7cm]{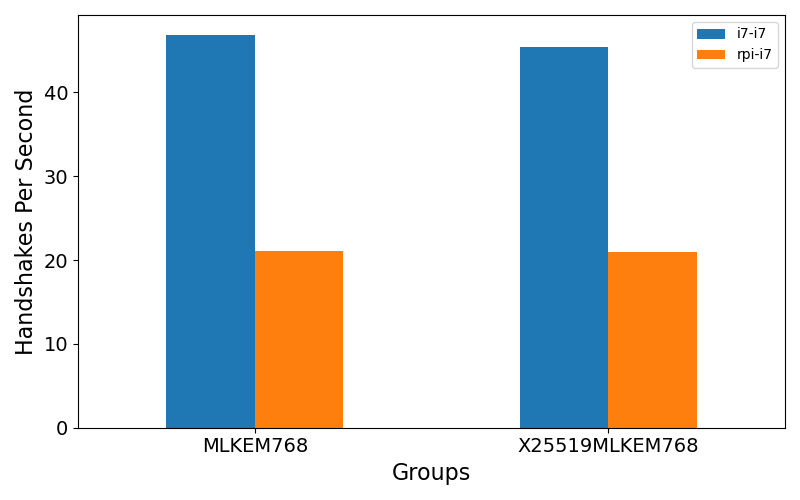}
    \caption{The \texttt{openssl s\_client} loop benchmark on an Intel i7 and a Raspberry~Pi~4.}
    \label{fig:9}
\end{figure}

The second stage evaluated the TLS~1.3 handshake performance results obtained using \texttt{openssl s\_time}, measured in terms of completed connections per second (CPS), across an Intel Core i7 workstation and a Raspberry~Pi~4, Fig.~\ref{fig:9} after following the workflow in Fig.~\ref{fig:1}c. The evaluation compares classical, post-quantum, and hybrid cryptographic configurations under identical experimental conditions.

Across both platforms in Fig.~\ref{fig:9}, classical signature algorithms such as ED25519 and ED448 consistently achieve the highest handshake throughput. This behavior reflects the maturity and computational efficiency of classical elliptic curve based signature schemes, which incur relatively low signing and verification costs during the TLS handshake. On the Intel i7 platform, classical configurations achieve CPS values exceeding those of all post-quantum and hybrid alternatives, while on the Raspberry~Pi~4 the same ordering is preserved, though at lower absolute throughput due to hardware constraints.

Post-quantum signature algorithms, including Falcon and ML-DSA parameter sets, exhibit lower handshake throughput compared to classical signatures but consistently outperform hybrid configurations. The reduced CPS observed for PQC signatures is attributable to higher computational complexity and larger key and signature sizes, which increase both cryptographic processing time and handshake message size. Falcon exhibits lower handshake throughput compared to ML-DSA, as it relies on floating-point arithmetic, which is costly to implement securely and efficiently in software. Nevertheless, these schemes remain viable in practice, particularly on high performance platforms, where the performance gap relative to classical signatures is reduced. 

Hybrid configurations combining classical key exchange (X25519) with ML-KEM-768 consistently show the lowest handshake throughput across all tested signature algorithms. This reduction arises from the additional cryptographic operations and increased handshake message sizes introduced by the hybrid key exchange, which must execute both classical and post-quantum key establishment mechanisms within a single handshake.

A direct comparison of key exchange groups further highlights this effect. For every signature algorithm evaluated, the classical X25519 key exchange achieves higher CPS than the hybrid X25519+ML-KEM-768 configuration. This trend holds across both platforms: on the Intel i7, classical key exchange consistently outperforms hybrid configurations, while on the Raspberry~Pi~4 the same relative ordering persists despite significantly reduced absolute throughput. Energy consumption was monitored using Intel RAPL package counters during benchmarking. Under identical conditions, X25519 consumed approximately 166.6~J (25.17 mJ/connection), while X25519+ML-KEM-768 consumed 201.7~J (34.33 mJ/connection), reflecting higher cumulative energy due to additional cryptographic processing.
These results demonstrate that, while hybrid mechanisms provide transitional quantum resistance, they introduce measurable performance penalties.

\begin{figure}[t]
    \centering
    \includegraphics[width=9.5cm]{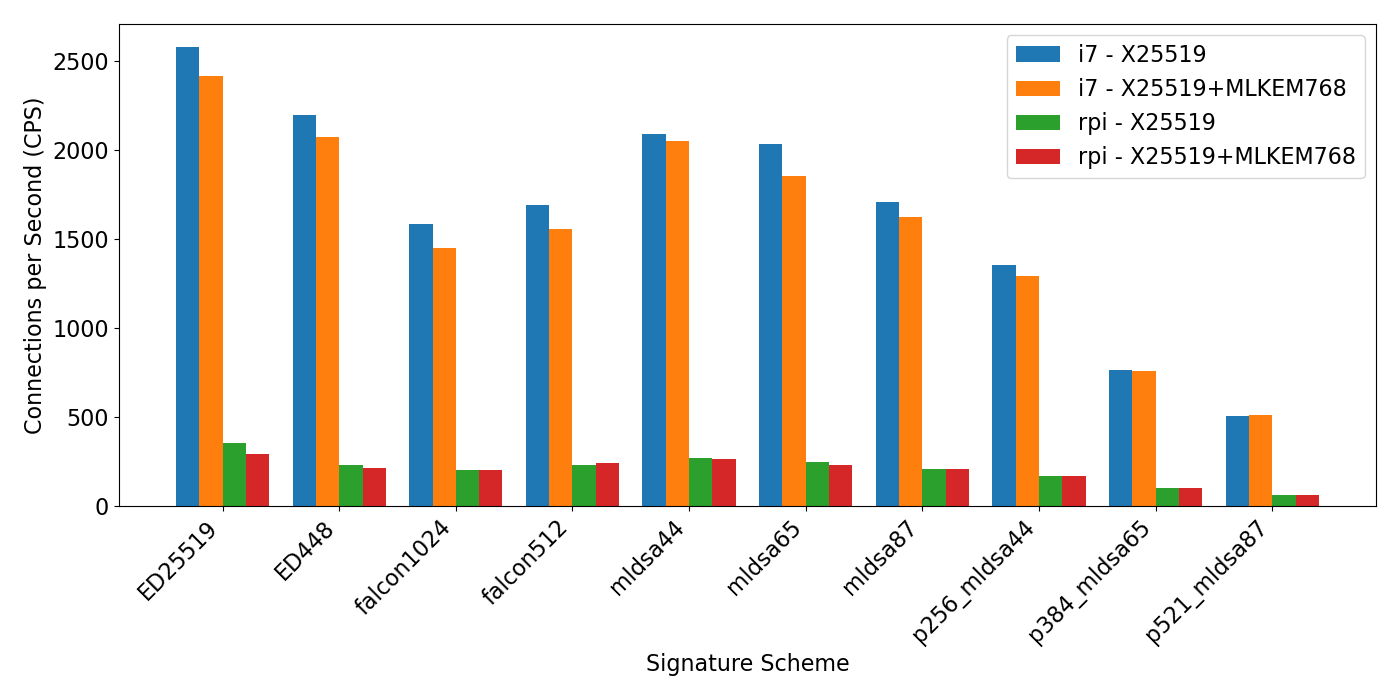}
    \caption{Evaluating \texttt{openssl s\_time} results on intel i7 and RaspberryPi4.}
    \label{fig:10}
\end{figure}

Fig.~\ref{fig:9} presents the TLS~1.3 handshake throughput, measured in connections per second (CPS), using the evaluation workflow illustrated in Fig.~\ref{fig:1}d. It compares ML-KEM-768 and X25519+ML-KEM-768 on two paths: i7--i7 and Raspberry~Pi--i7. The main observation is the clear throughput drop when moving from the homogeneous desktop-class path to the heterogeneous edge-style path. At the same time, ML-KEM-768 and X25519+ML-KEM-768 remain close in CPS on both paths, indicating that starting with a hybrid key exchange is unlikely to introduce a major throughput penalty during migration to PQC in this setup. This suggests that hybrid deployment can serve as a practical transitional step, allowing post-quantum mechanisms to be introduced without a large loss in TLS session-establishment rate. In contrast, prior TLS/X.509 benchmarking reported a much larger separation between classical and ML-KEM/hybrid throughput in localhost experiments~\cite{sim_kpqc_tls}.

Overall, based on full TLS 1.3 handshakes without session resumption, PQC-enabled TLS sustains throughput comparable to classical configurations under ideal, loss-free conditions. CPU utilization remained similar across X25519 (41\%), ML-KEM-768 (42\%), and X25519+ML-KEM-768 (44\%), with modest peak memory (\~10.7 MB), indicating that CPU saturation did not dominated performance on the i7 platform.\\
However, when artificial delay and packet loss were introduced using \texttt{tc netem}, throughput decreased sharply (e.g., 34 connections with 50 ms delay and 69 connections with 20 ms delay plus 0.5\% loss over 11 seconds from 5600 handshakes using falcon512), demonstrating that handshake scalability becomes strongly network-limited under realistic impairments.

\subsection{Ciphertext Compression and Expansion Rate}

Beyond computational performance, communication overhead is a defining constraint for PQC deployment in bandwidth- and latency-sensitive 6G architectures. Ciphertext expansion rate (CER) quantifies this overhead at the algorithmic level by measuring the ratio between transmitted ciphertext size and encapsulated payload. As shown in Table~\ref{tab:1}, CER increases with higher security levels and decreases with stronger compression, reflecting the trade-off between security strength, reliability, and communication efficiency.

In distributed 6G systems, the impact of CER becomes significant at the protocol layer. Security associations are repeatedly established across edge nodes and service-based interfaces, causing algorithm-level expansion to propagate directly into signaling traffic. As shown in Table~\ref{tab:2}, TLS handshake measurements confirm that increased ciphertext size results in measurable growth in control-plane payload. For example, ML-KEM-768 introduces approximately 2~kB of additional transmitted data per connection establishment, while PQC authentication significantly elevates the baseline signaling cost.

In highly distributed 6G topologies, this overhead accumulates across numerous short-lived or concurrent secure sessions. Larger handshake payloads increase the volume of data requiring transport during setup, raising fragmentation likelihood under constrained paths and proportionally increasing retransmission cost over wireless links. Consequently, ciphertext expansion influences control-plane scalability and connection robustness.

Therefore, CER should be interpreted not only as a cryptographic sizing metric (Table~\ref{tab:1}), but as a system-level scaling parameter whose effects are observable at the transport layer (Table~\ref{tab:2}).

\begin{table}
\centering
\caption{Ciphertext Expansion Rate (CER) for $K=256$ bits under different key exchange schemes and compression settings}
\label{tab:1}
\begin{tabular}{lccc}
\hline
\textbf{Scheme} & $k$ & $(d_u, d_v)$ & \textbf{CER} \\
\hline

ML-KEM-512 
& $2$ & $(12,12)$ 

& $36.0$ \\

ML-KEM-768 
& $3$ & $(12,12)$ 

& $48.0$ \\

ML-KEM-1024 
& $4$ & $(12,12)$ 

& $60.0$ \\

ML-KEM-512 (uniform comp.) 
& $2$ & $(10,4)$ 

& $24.0$ \\

ML-KEM-768 (uniform comp.) 
& $3$ & $(10,4)$ 

& $34.0$ \\

ML-KEM-1024 (uniform comp.) 
& $4$ & $(11,5)$ 

& $49.0$ \\
\hline
\end{tabular}
\end{table}

\begin{table}[t]
\centering
\caption{TCP Payload Bytes per TLS 1.3 Handshake for Different Algorithm Configurations}
\label{tab:2}
\begin{tabular}{lc}
\hline
\textbf{Scheme} & \makecell{\textbf{Handshake payload (bytes)}} \\
\hline
X25519 + RSA & 1893 \\
X25519 + MLDSA44 & 7257 \\
X25519MLKEM768 + RSA & 4157 \\
X25519MLKEM768 + MLDSA44 & 9521 \\
MLKEM512 + RSA & 3389 \\
MLKEM512 + MLDSA44 & 8753 \\
MLKEM768 + RSA & 4093 \\
MLKEM768 + MLDSA44 & 9457 \\
MLKEM1024 + RSA & 4957 \\
MLKEM1024 + MLDSA44 & 10321 \\
\hline
\end{tabular}
\end{table}

\section{Implications}
\subsection{Implications for 6G Architecture}

The results indicate that PQC integration in future 6G networks is strongly context-dependent. In core network environments with abundant computational resources and long-lived security associations, ML-KEM-based key exchange introduces manageable overhead. However, in edge and IoT scenarios, ciphertext expansion becomes a dominant constraint, necessitating compression-aware design and careful parameter selection. For latency-critical URLLC services, frequent full TLS handshakes can further amplify computational and bandwidth costs. Recent large-scale deployments have begun addressing handshake latency by proactively scanning origin capabilities and directly transmitting the preferred hybrid keyshare in the initial ClientHello, thereby eliminating unnecessary HelloRetryRequest round trips and improving both security and performance \cite{cloudflare_automatically_secure}.

\subsection{Implications for Standards}
The adoption of PQC in telecom networks implies that security standards must evolve from algorithm selection to system-level integration and migration guidance. PQC mechanisms are not drop-in replacements for classical public-key cryptography; they introduce larger key sizes, higher computational cost, and different handshake semantics, all of which must be explicitly addressed in standards.

For Third Generation Partnership Project (3GPP), this implies that future releases must specify PQC-ready security options for core procedures such as authentication and key agreement, subscriber identity protection, and inter-network trust. Standards will need to define how PQC and hybrid schemes are negotiated, how backward compatibility is maintained, and how performance constraints in radio and core networks are handled \cite{marin_pqc_telecom_2024}.

For European Telecommunications Standards Institute (ETSI), the implication is a stronger focus on migration frameworks and hybrid cryptographic profiles, enabling operators to deploy quantum-safe mechanisms incrementally while maintaining interoperability with legacy systems. ETSI specifications are likely to shape practical deployment guidance and regulatory compliance requirements \cite{marin_pqc_telecom_2024}.

For the GSM Association (GSMA), the implication is the need to translate cryptographic standards into operator focused guidelines, addressing real world issues such as roaming between heterogeneous networks, certificate management during transition, and operational risk. GSMA’s role as a liaison also implies increasing alignment between operator requirements and formal standardization \cite{marin_pqc_telecom_2024}.

At the global level, NIST’s PQC standards imply a convergence toward a common set of cryptographic primitives across telecom, cloud, and government systems, reducing fragmentation but increasing the importance of efficient implementations. Similarly, IETF standardization of PQC in Internet protocols implies that telecom networks must ensure consistency between telecom specific security frameworks and Internet-based protocols used in cloud-native network functions \cite{marin_pqc_telecom_2024}.

Overall, PQC adoption in telecom must be standards-driven, performance-aware, and migration-focused. Standards must address not only cryptographic correctness, but also deployment constraints, interoperability, and long-term operability across core, edge, and IoT domains.

\section{Opportunities: Optimizing PQC for 6G}

The evaluation highlights several opportunities to improve PQC efficiency for bandwidth and energy constrained 6G deployments, particularly by reducing ciphertext expansion (Table~\ref{tab:1}) while preserving reliability.

\begin{itemize}

\item {\emph{Standardized compression refinement:}}
ML-KEM employs uniform coefficient compression as defined in NIST FIPS~203~\cite{nist_fips203}, ensuring a bounded decryption failure rate (DFR). While this baseline preserves reliability, further optimization of compression parameters beyond the standardized bounds can reduce signaling overhead but may compromise correctness guarantees.

\item {\emph{Non-standard quantization techniques:}}
Lloyd–Max and semi-compressed approaches~\cite{liu_sc_kyber_2025} modify the uniform scalar compression defined in FIPS~203 by introducing noise-aware, non-uniform quantization or selectively compressing ciphertext components after symbol mapping (e.g., only the $u$ polynomial). By better matching the underlying lattice noise distribution, these methods can reduce mean-squared quantization error and slightly lower DFR at similar communication overhead. However, this improvement comes at the cost of higher computational complexity and increased implementation effort, which must be carefully evaluated for constrained 6G deployments.

\item {\emph{Advanced lattice-based compression:}}
Structured vector quantization methods (e.g., Barnes–Wall or E8 constructions) can reduce effective ciphertext expansion compared to per-coefficient truncation~\cite{liu_lattice_quantizer_2024}. Such techniques increase implementation complexity and may introduce side-channel vulnerabilities.

\item {\emph{Symbol mapping with dual error correction:}}
Mapping coefficients to higher-order constellations (e.g., 8-PAM) increases bit density per symbol. When combined with Gray coding and lightweight dual-layer error correction, both bit-level and burst errors can be mitigated under stronger compression. Any decoding and correction logic must be constant-time to prevent timing leakage.

\item {\emph{Hardware acceleration:}}
Dedicated support for polynomial arithmetic and modular reduction can reduce latency and energy consumption without altering standardized cryptographic parameters.

\end{itemize}

Optimizing PQC for 6G requires balancing ciphertext expansion, reliability, computational cost, and security. While FIPS 203 guarantees bounded DFR, more aggressive compression can lower communication overhead at the cost of added complexity and potential reliability trade-offs. Any optimization must maintain telecom-grade DFR, security, and standards interoperability.

\section{Open Challenges and Research Gaps}

Despite PQC standardization, its integration into 6G remains a systems-level challenge.

\begin{itemize}

\item {\emph{Lack of 6G-specific deployment profiles:}}
There is no standardized mapping between PQC parameter sets and distinct 6G service classes (core, edge, URLLC), leaving bandwidth and latency thresholds undefined~\cite{ericsson_mlkem_missing}.

\item {\emph{Limited energy-aware evaluation:}}
Energy consumption and memory footprint remain insufficiently characterized for battery-powered edge and IoT deployments.

\item {\emph{Interoperability complexity:}}
Hybrid operation across classical, post-quantum, and emerging quantum communication mechanisms introduces migration, negotiation, and compatibility challenges.

\item {\emph{Reliability under wireless conditions:}}
The behavior of compressed PQC schemes under fading and retransmissions remains insufficiently characterized. While standardized parameters bound DFR, aggressive compression or non-standard configurations may introduce reliability, interoperability, and compliance risks. Any optimization must therefore preserve FIPS-compliant parameters, security, and cross-vendor compatibility for safe 6G deployment.

\item {\emph{Limited end-to-end validation:}}
The absence of mature 6G testbeds restricts cross-layer evaluation from cryptographic primitive to radio-layer signaling.

\end{itemize}

\section{Conclusion}

This work shows that PQC can be integrated into network infrastructures with manageable overhead, while highlighting ciphertext expansion and deployment-level evaluation as important challenges for 6G. Algorithm-level benchmarking, TLS handshake measurements, and architectural analysis indicate that PQC is feasible for next-generation connectivity, but requires careful parameter selection.

Looking ahead, combining lattice-based schemes such as ML-KEM with physical-layer approaches like QKD can support long-term secure and resilient 6G networks.

\bibliographystyle{IEEEtran}

\bibliography{IEEEabrv,references}

\vspace{-3\baselineskip}
\begin{IEEEbiographynophoto}{Ananya Kudaloor}
is a Research Engineer with the Bristol Research and Innovation Laboratory of Toshiba. Her work focuses on analysis, optimization, and implementation of PQC techniques. 
\end{IEEEbiographynophoto}
\vspace{-3\baselineskip}
\begin{IEEEbiographynophoto}{Adnan Aijaz} currently holds the positions of Chief Research Fellow and Programme Leader at the Bristol Research and Innovation Laboratory of Toshiba. His recent activities focus on industrial systems, 5G/6G, Open RAN, and Energy/Quantum Internets. 
\end{IEEEbiographynophoto}
\end{document}